\documentclass[twocolumn]{aastex631}

\usepackage{soul}
\usepackage{slashbox}
\usepackage[version=4]{mhchem}

\shorttitle{Photochemistry induced spectral features}
\shortauthors{Locci et al.}

\graphicspath{{./}{figures/}}

\begin{document}

\title{Signatures of X-ray dominated chemistry in the spectra of exoplanetary atmospheres}

\author[0000-0002-9824-2336]{Daniele Locci}
\affiliation{INAF - Osservatorio Astronomico di Palermo, P.za Parlamento 1, 90134 Palermo, Italy}

\author[0000-0003-3683-0834]{Giambattista Aresu}
\affiliation{INAF - Osservatorio Astronomico di Cagliari, Via della Scienza 5, 09047 Selargius, Italy}

\author[0000-0002-9882-1020]{Antonino Petralia}
\affiliation{INAF - Osservatorio Astronomico di Palermo, P.za Parlamento 1, 90134 Palermo, Italy}

\author[0000-0002-9900-4751]{Giuseppina Micela}
\affiliation{INAF - Osservatorio Astronomico di Palermo, P.za Parlamento 1, 90134 Palermo, Italy}

\author[0000-0001-5154-6108]{Antonio Maggio}
\affiliation{INAF - Osservatorio Astronomico di Palermo, P.za Parlamento 1, 90134 Palermo, Italy}

\author[0000-0001-7480-0324]{Cesare Cecchi-Pestellini}
\affiliation{INAF - Osservatorio Astronomico di Palermo, P.za Parlamento 1, 90134 Palermo, Italy}

\begin{abstract}
High-energy radiation from stars impacts planetary atmospheres deeply affecting their chemistry, providing departures from chemical equilibrium. While the upper atmospheric layers are dominated by ionizations induced by extreme ultraviolet radiation, deeper into the atmosphere molecular abundances are controlled by a characteristic X-ray dominated chemistry, mainly driven by an energetic secondary electron cascade. 
In this work, we aim at identifying molecular photochemically induced fingerprints in the transmission spectra of a giant planet atmosphere. We have developed a numerical code capable of synthesizing transmission spectra with arbitrary spectral resolution, exploiting updated infrared photoabsorption cross sections. Chemical mixing ratios are computed using a photochemical model, tailored to investigate high energy ionization processes. 
We find that in case of high levels of stellar activity, synthetic spectra in both low and high resolutions show significant, potentially observable  out-of-equilibrium signatures arising mainly from CO, CH$_4$, C$_2$H$_2$, and HCN.
\end{abstract}


\submitjournal{Planetary Science Journal}

\section{Introduction}
The formation and evolution of planets is strictly linked to the evolution of their parent stars. The many processes in star-planet interaction play a critical role in shaping the great diversity of planetary atmospheres, either in the solar- and exo-systems. While the host star’s optical and infrared radiation is the primary driver of thermal structure of the lowest atmospheric regions, the high energy portion of the stellar spectrum, the extreme ultraviolet (EUV) radiation and the X-rays  emitted mainly by the corona, may modify the physical structure of the atmosphere. Radiation in such energy bands may heat the outer layers, leading to a phase of hydrodynamic instability (e.g.~\citealt{Murray-Clay09,Owen19}), or cause a vigorous photochemistry (e.g.,~\citealt{Garcia07,Moses11,Locci22}) providing a route to disequilibrium. 

This high energy portion of the stellar spectrum (also know as XUV radiation) is variable in time, and related to the stellar activity and evolution (e.g.,~\citealt{Ribas05}). Young stars are typically in the saturation phase, during which the X-rays luminosity, $L_{\rm X}$, depends only on the stellar mass and age, following the bolometric luminosity in a roughly constant ratio $L_{\rm X}/L_{\rm bol}=10^{-3}$ \citep{Pizzolato03}. A star stays in the saturation phase for a period of time that depends on the initial mass and rotation state.  F and G stars have larger $L_{\rm bol}$  than M stars, and consequently larger X-ray luminosities; lower mass stars generally remain in the saturation phase for longer times than massive stars, ending up with a larger time-integrated XUV flux \citep{Johnstone21}. After the saturation phase, the X-ray luminosity follows the fate of the stellar rotation rate that decreases in time because of magnetic braking. This causes the X-rays luminosity to decrease, at a pace that may depend on the energy range (e.g., \citealt{Micela02}). As a consequence, planets in close orbits around stars of moderate mass are more subjected to intense hydrodynamic escape, while  planets around young and/or massive stars are expected to present an active photochemistry.

Precise measurements of chemical abundances in planetary atmospheres are necessary to constrain the formation histories of exoplanets. High-resolution transmission spectroscopy is a powerful (and successful) technique in detecting molecular features in transiting exoplanetary atmospheres (e.g., \citealt{Giacobbe21}). In this work we derive synthetic absorption spectra of planetary atmospheres of  gaseous giants, irradiated by a broad band XUV spectrum. Such calculations are performed with a radiative transfer code tailored for modeling the  photochemistry of primordial type atmospheres dominated by hydrogen and helium. 

As shown in \citet{Locci22} the upper  layers in these environments ($p \lesssim 10^{-6}$~bar) are dominated by primary ionization provided by EUV radiation, and are thus, mainly populated by atomic and ionic species (e.g., \citealt{Chadney16}). X-rays penetrate deeper in  an atmosphere with with solar-like abundances \citep{CCP06,CCP09}, because of the small absorption cross-sections of major atmospheric  constituents. The most energetic photons may thus reach pressures as high as $p \sim 10^{-3}-10^{-2}$~bar. Fast electrons produced in primary ionization events have enough  energy to produce further ionizations. In the X-ray energy range, primary electrons give rise to a secondary cascade that dominate the energy deposition throughout the illuminated regions of the atmosphere, with the exception of the uppermost layers. The total ionization rate of the medium is given by the sum of primary and secondary ionization rates (e.g., \citealt{Maloney96}). Unlike X-ray photons, primary electrons are able to efficiently ionize molecular hydrogen, giving rise to a characteristic chemistry, rich in molecular ions such as \ce{H3O+} and \ce{NH4+} \citep{Locci22}. A further notable consequence of secondary ionizations is the significant increase in the abundances of species like \ce{CH4}, \ce{C2H2}, and \ce{HCN} \citep{Locci22}, that may significantly contribute to atmospheric spectra.

Frequently, atmospheric transit detections are described by wavelength-dependent variations in the apparent planetary radius through a variety of photometric and spectroscopic techniques (e.g.,~\citealt{Seager10}). Such methods exploit both high- and low-resolution data, or their combinations. The advent 

The advent of \emph{JWST} (James Webb Space Telescope) and the future launch of \emph{ARIEL} (Atmospheric Remote-sensing Infrared Exoplanet Large-survey space) mission \citep{Tinetti18, Tinetti21} is providing a wealth of data to characterize planetary atmospheres, formed and evolved under different physical and chemical conditions. This requires the production of spectroscopic data and a better insight into chemical reaction kinetics, through which provide the synthesis of planetary spectra at low and high resolutions. In this work we show how an accurate modelling of radiation transfer and photochemistry (including ionization processes) allows for the production of high resolution synthetic spectra, capable to discriminate between equilibrium and out of equilibrium chemistries.  This way, the high altitude locations and low pressures where radiation-induced disequilibrium processes take place could be, in principle, determined. In Section~\ref{sez:chem} we describe briefly the chemical code we use to construct abundance profiles. The cross-section database needed for deriving the transmission spectra is presented in Sections~\ref{specDB} and~\ref{sez:trasm}. Results are shown  in Section~\ref{sez:ris}, and discussed in Section~\ref{sez:dis}, where we also draw our conclusions. 

\section{Chemistry} \label{sez:chem}
Abundance profiles within the atmosphere are derived using the chemical kinetics code described in~\cite{Locci22}, that calculates the gas phase distribution of about 130 species, in a gas of solar-like composition, comprising hydrogen, helium, oxygen, carbon, and nitrogen. The model solves the continuity equation for 1D stratified spherical atmospheres, and describes accurately the energy deposition of ionizing radiation, as reported in~\cite{Locci18}. The secondary electron cascade induced by primary photo-ionization is computed exploiting the Monte Carlo technique described in ~\cite{CCP06}. A key parameter in the model is the mean energy per ion pair, i.e., the ratio of the primary photo-electron energy to the number of ionizations provided before the primary electron comes to rest \citep{Dalgarno99}. 

In photochemical models,  the incident stellar spectrum is (by definition) a critical factor. In our model we exploit a mosaic from several sources: the ultraviolet spectral band is taken from the library \emph{PHOENIX} \citep{Husser13}, to  which we add a Lyman-$\alpha$ profile;  XUV radiation encompass the range between the ionization potential of atomic hydrogen (13.6 eV) to 10 kEV; the EUV portion,  extending up to 100~eV, assumes a constant value, with an integrated luminosity $L_{\rm EUV}$; finally, the X-ray spectrum is described by the emission of an optically thin plasma \citep{Raymond77}, with an integrated luminosity $L_{\rm X}$. The total luminosities of the Lyman-$\alpha$ line and $L_{\rm EUV}$ are scaled from the X-ray luminosity by means of the prescriptions given in \citet{Linsky20} and \citet{Sanz-Forcada11}, respectively.  We note that different X-ray to EUV scaling relations have been recently put forward (e.g., \citealt{Chadney15, Johnstone21})., that may modify the luminosity of the EUV component, typically providing EUV luminosities lower than the one provided by \citet{Sanz-Forcada11} relation. However, the effects of EUV are important mainly in the outermost layers, and furthermore, EUV photons interact preferentially with hydrogen and helium. As a consequence, the impact on chemistry is confined only to very low pressures, while the layers probed by infrared spectroscopy are virtually unaffected.

\section{Building the spectral database}\label{specDB}
We build high resolution spectra ($R \sim 100,000$) for the species listed in Table \ref{SP_tab}. For all the species we used the Exomol Database (see \citealt{Tenn2020} and references therein): we downloaded the \textit{.states} and \textit{.trans} (transitions) files and generated the stick spectra, which constitute the input files for the spectral code.

The Exomol transitions files contain many molecular transitions and are generally of considerable size, reaching billions of lines for molecules such as \ce{H2O}. To optimize line by line computation it is desirable to cutoff weaker lines when building the stick spectra from scratch. This is done following \cite{Rothman13}, which uses the following criteria for the intensity cutoff 
\begin{eqnarray}
I_{\rm lim} &=& S_{\rm cr}\left( \frac{\nu}{\nu_{\rm cr}} \right) {\rm tanh} \left( \frac{hc\nu}{2k_{\rm B}T} \right) \quad \nu \le \nu_{\rm cr}\nonumber\\
I_{\rm lim} &=& S_{\rm cr} \quad \nu > \nu_{\rm cr}\nonumber
\end{eqnarray}
where $S_{\rm cr}$ is the critical line intensity, $\nu_{\rm cr}$ the critical wave number, $T$ is the temperature at which the calculation is done. We adopt $S_{\rm cr} = 10^{-30}$~cm/molecule and $\nu_{\rm cr}=2000$~cm$^{-1}$ for all the molecules used here apart from \ce{CH4}. Given the very high number of lines in the wavelength range where the resolution is the highest, for methane we chose $S_{\rm cr} = 10^{-28}$ cm/molecule. The critical frequency cutoff ensures that more transitions are allowed in a low frequency spectral region  where lines are less dense. We may evaluate these criteria at  arbitrary (relevant) temperatures, and keep all the transitions that satisfy the cutoff at least once. In this paper, we include spectra calculated at 1000 K,  i.e., the assumed temperature of the atmosphere (see Section \ref{sez:ris}).  In the HITRAN database, the intensity cut-off is generally set to $1 \times 10^{-29}$~cm/molec (e.g., \citealt{Yur18} and references therein). We select a an intensity cut-off a factor 10 lower. Even this choice does not provide more than a small percent of the \ce{H2O} and \ce{CO} lines available in the database. However, most of the unselected transitions have very low intrinsic intensity, and whose addition would produce very small (if any) effects on the final spectra. This is the criterion, we  exploit also for methane. Simulation suggest that an intensity cut-off  $1 \times 10^{-28}$~cm/molecule would not produce a substantial effect on reference spectra. This occurs in particular for those bands that emerge over the water spectrum, being them formed by the strongest lines in the database. The increase of the cut-off intensity up to $1 \times 10^{-30}$~cm/molecule does not provide changes in the quality of the \ce{CH4} spectra presented in this study.

We validated our spectral code against EXOCROSS \citep{Yur18} for the \ce{H2O} spectra between 1000 and 5000 cm$^{-1}$: the relative difference between the two spectra is of the order of 0.2\% on average, always contained well within 0.5\%. In column 4 of Table \ref{SP_tab} we indicate the fraction of lines retained compared to the complete Exomol database after applying the line cut off described above.

\begin{table}[h]
\caption{List of molecules for which the spectra was calculated. The percentage of line retained for the actual calculation depends on the line intensity cutoff criteria discussed in section \ref{specDB}.}.
\centering
\begin{tabular}{l|c|c|c|c}
Molecule  & lines & retained (\%) & Intensity cutoff\\
\hline
\ce{H2O}  & $5.7\times 10^9$ & $\sim 5$ & 10$^{-30}$\\
\ce{CO}   & $1.2\times 10^5$ & 100 & -\\
\ce{CO2}  & $2.6\times 10^9$ & $\sim 4$ & 10$^{-30}$\\
\ce{NH3}  & $1.7\times 10^{10}$ & $\sim$38 & 10$^{-30}$\\
\ce{HCN}    & $3.4\times 10^7$ & $100$ & - \\
\ce{C2H2} & $4.3\times 10^9$ & $\sim$86 & 10$^{-30}$ \\
\ce{CH4} & $4.3\times 10^9$ & $\sim$74 & 10$^{-28}$ \\
\ce{H2}    & $4.7\times 10^3$ & 100 & - \\
\hline
\end{tabular}
\label{SP_tab}
\end{table}

The line shape is a Voigt profile, the calculation of which is extended for each line out to 25 cm$^{-1}$ from the line center \citep{Yur18}. The stick spectra is used by the spectral code to build the absorption spectra, making use of the Exomol \textit{.broad} files (to account for the collisional broadening with H$_2$, He or the species itself) and of the \textit{.shift} files (to account for the pressure shift). 

\section{Building the transmission spectra}\label{sez:trasm}
We calculate the transmission spectra assuming a 2D geometry, in which one of the axis is directed along the line joining the planet center to the substellar point (defining the horizontal direction). The values of altitude ($h$), pressure ($p$), mixing ratio ($w$), and number density ($n$), are assigned to any grid point located at least at one planetary radius $R_{\rm P}$ from the center. At the planetary radius, we set $h = 0$ and the maximum pressure $p_0 = 1$~bar. The pressure is assumed to decrease upwards. When $h < 0$ we consider the stellar radiation to be fully absorbed by the atmosphere. For each horizontal line of sight of vertical coordinate $h = h_k$, we calculate the optical depth as follows
\begin{equation}
\displaystyle \tau(\lambda,h_k) = {2 \sum_i \int_0^{x_f(h_k)} \hspace{-0.4cm} \sigma_i(\lambda)n_i(x,h_k) dx}
\end{equation} 
where $\sigma (\lambda)$ is the absorption cross-section of the $i-$th species (see Section~\ref{specDB} and Table~\ref{SP_tab}), $x_f(h_k)$ is the horizontal coordinate of the last atmospheric point at altitude $h = h_k$, and the factor 2 accounts for the symmetrical hemisphere. We convolve the contributions of all sightlines calculating the equivalent depth (defined in \citealt{Hollis13}) as
\begin{equation}
\displaystyle A (\lambda)= 2 \int_{0}^{h_f} h \left[1-e^{-\tau(\lambda,h)} \right] dh
\end{equation} 
where $h_f$ is the maximum altitude of our simulation corresponding to a pressure $p = 10^{-10}$ bar. Given the spherical geometry, such final value of the pressure is replicated along all the lines of sight at $x = x_f$. Here, the factor two accounts for the southern hemisphere. The total transit depth is given by
\begin{equation}
D(\lambda)= \frac{R^2_{\rm P}+A(\lambda)} {R^2_\star}
\end{equation}
$R_\star$ being the stellar radius. In order to simulate the instrumental response, we apply to the transit depth a Gaussian filter to obtain
\begin{equation}
F(\lambda)= \int_{-\infty}^\infty D(\lambda^\prime)g(\lambda-\lambda^\prime)d\lambda^\prime
\end{equation}
where $g$ is a normalized Gaussian function truncated at $4 \sigma$,  $\sigma$ being the FWHM whose value depends on the required resolution. Finally, we calculate the contribution function 
\begin{equation}
c (\lambda,p > P)= \frac{A[\lambda,\tau(\lambda,p > P)=0]} {A(\lambda)}   
\label{conf}
\end{equation}
where $A[\lambda,\tau(\lambda,p > P)=0]$ is the equivalent atmospheric depth obtained setting to zero the optical thickness of those layers for which the pressure is larger than an assigned boundary value $P$. Finally, the integrated contribution function reads as
\begin{equation}
q (p > P) = \int_{\Delta \lambda}  c (\lambda,p > P)  d \lambda
\label{confi}
\end{equation}
with $\Delta \lambda$ the wavelength range in which the integration is performed. 

\section{Results}\label{sez:ris}
Using the photochemical code described in \citet{Locci22} and the prescriptions given in previous Sections, we synthesize a spectral distribution for each specific model representation. Our aim is to extract  XUV-driven features, and  evaluate their detectability, through either low and high resolution spectroscopy. We shall show computed spectra in the $1-10$~$\mu$m range, covered by a few instruments on board \emph{JWST}, the future planetary mission \emph{ARIEL}, and also accessible to ground based, high resolution telescopes. To this aim, we took into account a synthetic gaseous giant around a solar-like star. Parameters describing the star/planet system are reported in Table~\ref{tab:refmod}. The zenith angle adopted in the calculation of the chemical profiles is chosen to be $\theta = 60^\circ$, considered to be a good approximation for the globally averaged profile \citep{Johnstone18}. Spectra are derived for three different values of the resolving power, $R = 300$, 3,000, and 50,000. We also assume three different quiescent X-ray luminosity corresponding to as many stellar characteristics: $L_{\rm X} = 10^{26}$~erg s$^{-1}$, a rather  quiet star, $L_{\rm X} = 10^{28}$~erg s$^{-1}$, a star slightly more active than the Sun, and $L_{\rm X} = 10 ^{30}$~erg s$^{-1}$, a young active star.  X-ray luminosities are characterized by increasing plasma temperatures, namely $T_{\rm X} = 0.3$, 0.5 and 1~keV, respectively. For each model, we adopt a uniform temperature throughout the atmosphere $T = T_{\rm eq}=1000$~K, that also sets the planet orbital distance. 
\begin{table} 
\caption{Parameters of the star/planet system.}
\centering
\begin{tabular}{lll}
& parameter  & value \\
\hline
$\bullet$ & planet mass, $M_{\rm P}$    &  150 $M_\oplus$ \\
$\bullet$&planet radius, $R_{\rm P}$ &  12 $R_\oplus$\\
$\bullet$&planet equilibrium & 1000 K \\
&temperature, $T_{\rm eq}$ &  \\
$\bullet$&zenit angle, $\theta$ & 60° \\
$\bullet$& stellar X-rays & 10$^{26}$, 10$^{28}$, 10$^{30}$ erg s$^{-1}$  \\
& luminosity, $L_{\rm X}$ &  \\
\hline
\end{tabular}
\label{tab:refmod}
\end{table}

\subsection{Low resolution spectra}
As described  in \citet{Locci22}, the present photochemical model predicts significant enhancements in the abundances of some molecular species, following the X-ray induced chemistry.  Chemical effects are not solely  productive, as strong X-ray irradiation lowers appreciably the upper boundary of the residing regions of abundant species, such as e.g., water, carbon monoxide and dioxide. The response of species sensitive to high energy irradiation provides modifications of the atmospheric spectra, thus reflecting such non-equilibrium chemistry. 

\begin{figure}
\vspace{1cm}
\centering
\includegraphics[width=0.52\textwidth]{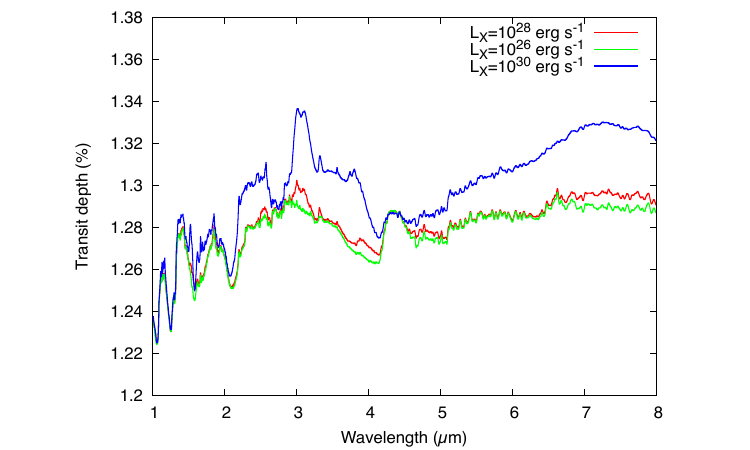}
\caption{Transmission spectra computed by varying the value of the X-rays luminosity: $L_{\rm X}=10^{26}$  (green line),  $10^{28}$ (red line), and $10^{30}$~erg~s$^{-1}$ (blue line),  with $T_{\rm X} = 0.3$, 0.5, and 1~keV, respectively.}
\label{fig:spLX}
\end{figure}

\begin{figure*}
\begin{tabular}{cc} 
\hspace{-1.1cm}\includegraphics[width=9cm]{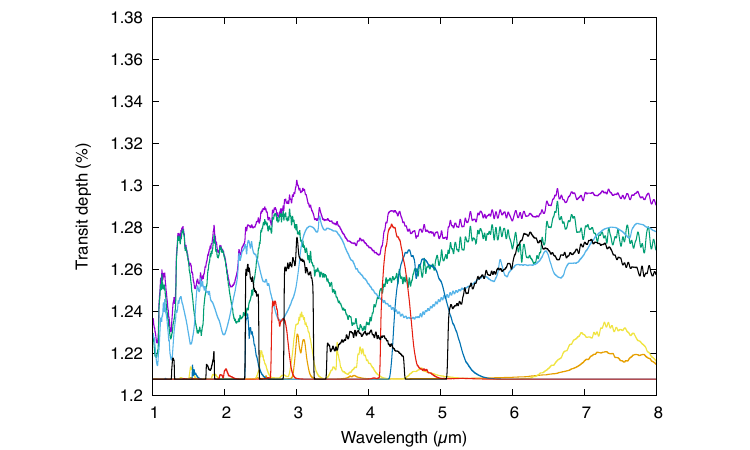} & \hspace{-0.5cm} \includegraphics[width=9cm]{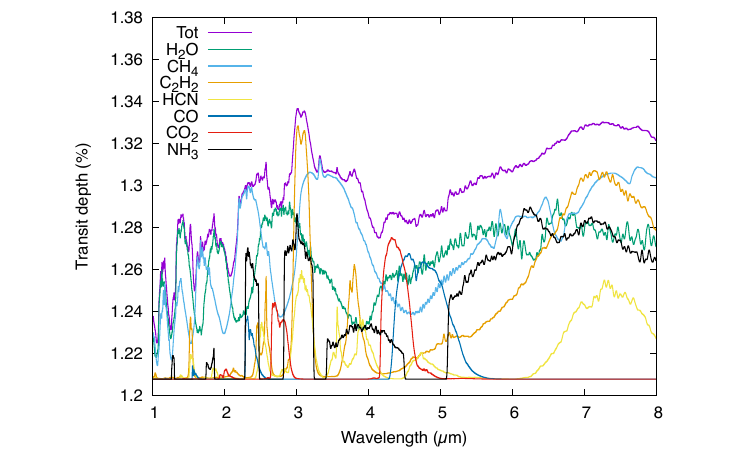}\\
\end{tabular}
\caption{Transmission spectra for $L_{\rm X} = 10^{28}$~erg s$^{-1}$ (left panel), and $L_{\rm X} = 10^{30}$~erg s$^{-1}$  (right panel), in which the contributions arising from individual species are highlighted.}
\label{fig:contributi}
\end{figure*}

We first examine the case of low-resolution spectra, $R = 300$ (Figure~\ref{fig:spLX}), for our three fiducial values of the X-ray luminosity. The resulting spectra indicate that the illuminating stellar radiation has a modest impact for low to moderate X-ray luminosities, with the spectra dominated by water features, superposed to strong \ce{CH4}  (around $3-4~\mu$m), and \ce{CO} and  \ce{CO2} ($4.2 - 5~\mu$m) signatures.
Increasing the luminosity to $L_{\rm X}= 10^{30}$~erg~s$^{-1}$, i.e., enhancing the stellar forcing,  intense features appear in a few bands between 3 and 4~$\mu$m, and beyond 5~$\mu$m. Specifically, the feature at around 3.1~$\mu$m is due to \ce{C2H2} and \ce{HCN}, the feature at $\sim 3.2-3.4~\mu$m to \ce{CH4}, the feature at around $4~\mu$m to \ce{HCN}, and those longwards 5~$\mu$m are mainly contributed by \ce{C2H2},  \ce{HCN}, \ce{CH4} and \ce{NH3}, as it is evident examining the contributions from individual species shown in Figure~\ref{fig:contributi}.

\begin{figure*}
\begin{tabular}{cc} 
\includegraphics[width=9cm]{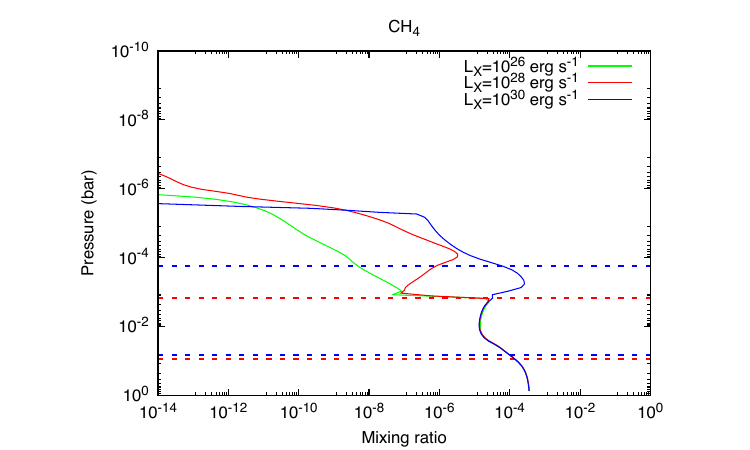} & \hspace{-1cm} \includegraphics[width=9cm]{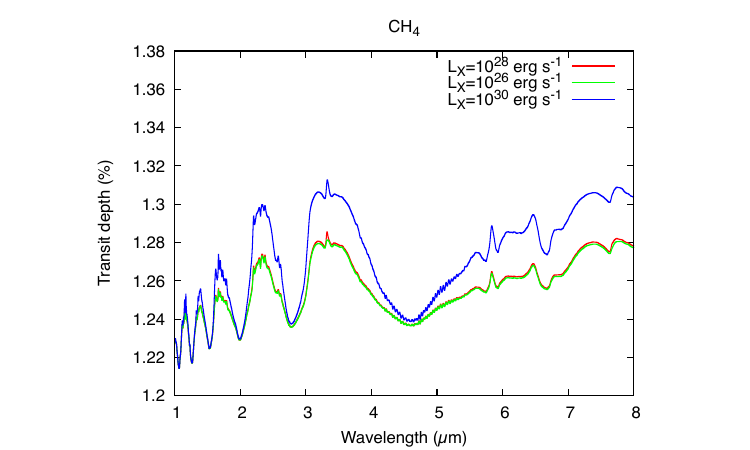}\\
\includegraphics[width=9cm]{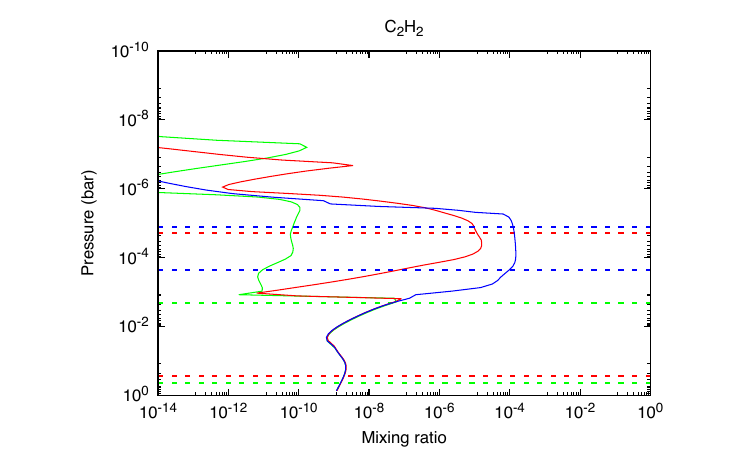} & \hspace{-1cm} \includegraphics[width=9cm]{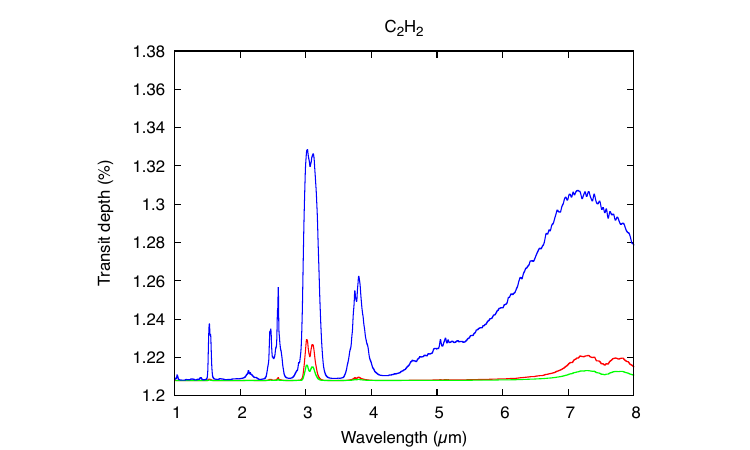}\\
\includegraphics[width=9cm]{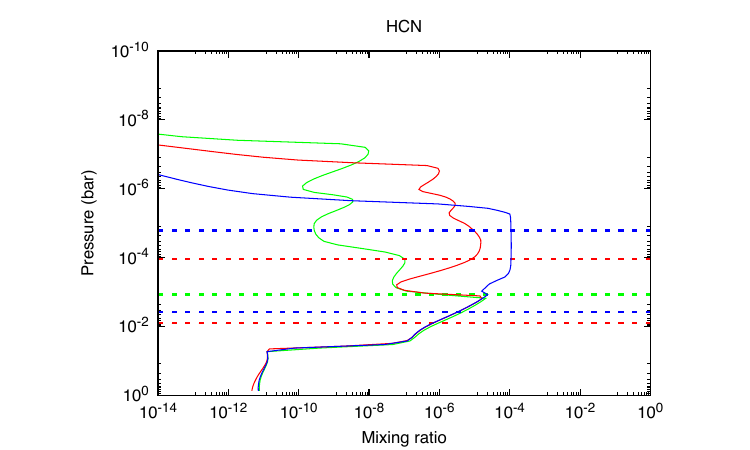} & \hspace{-1cm} \includegraphics[width=9cm]{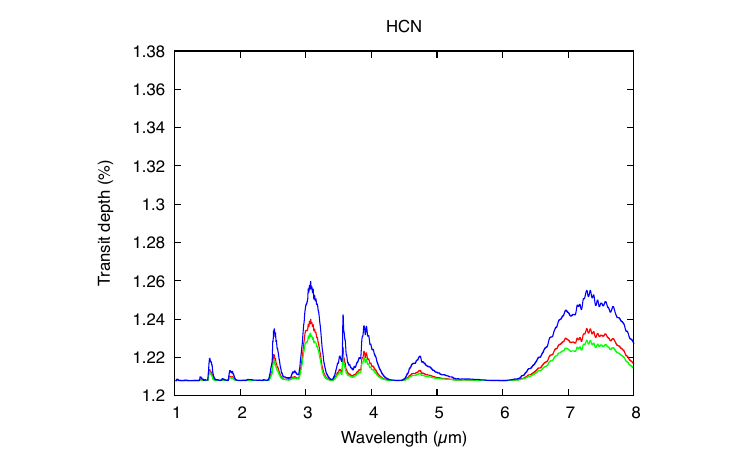}\\
\end{tabular}
\caption{Chemical vertical profiles for the species \ce{CH4}, \ce{C2H2} and \ce{HCN} positively reacting to the presence of XUV radiation (left panels), and their individual transmission spectra (right panels). Dashed lines in the left panels indicate  the pressure at which the integrated contribution function, equation (\ref{confi}), of the individual species is $q = 0.1$ and 0.9, for the 3 values of $L_{\rm X}$. Green lines: $L_{\rm X}=10^{26}$ erg s$^{-1}$; red lines: $L_{\rm X}=10^{28}$ erg s$^{-1}$; blue lines $L_{\rm X}=10^{30}$ erg s$^{-1}$.} 
\label{fig:abb1}
\end{figure*}

\begin{figure*}
\begin{tabular}{cc} 
\includegraphics[width=9cm]{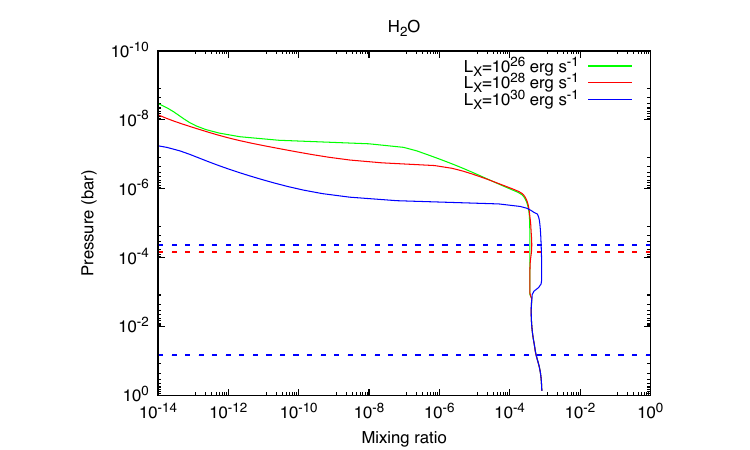} & \hspace{-1cm} \includegraphics[width=9cm]{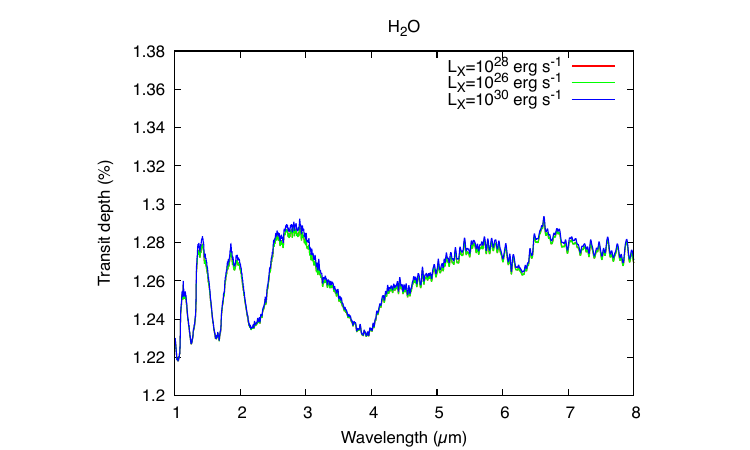}\\
\includegraphics[width=9cm]{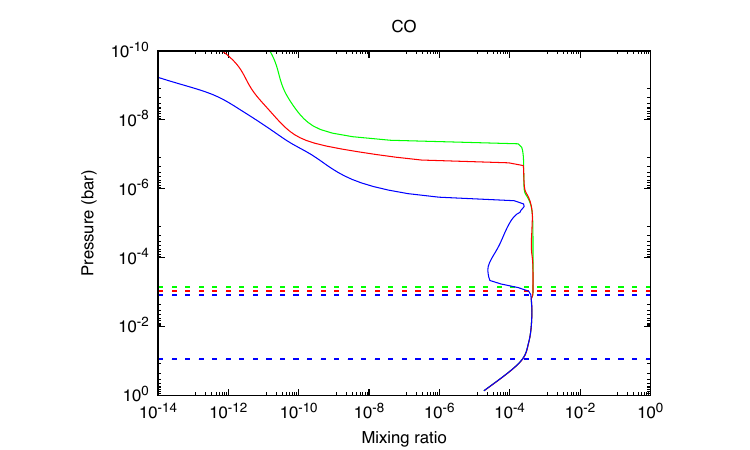} & \hspace{-1cm} \includegraphics[width=9cm]{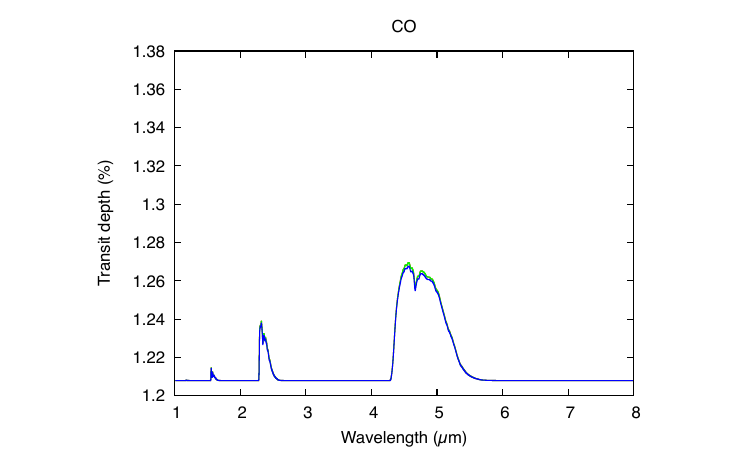}\\
\includegraphics[width=9cm]{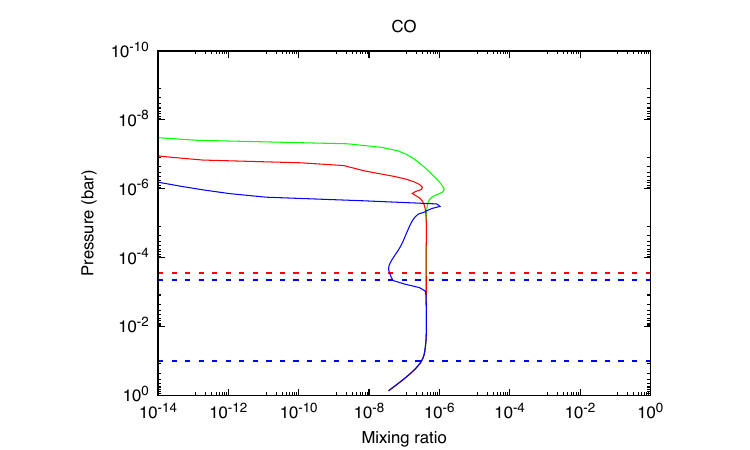} & \hspace{-1cm} \includegraphics[width=9cm]{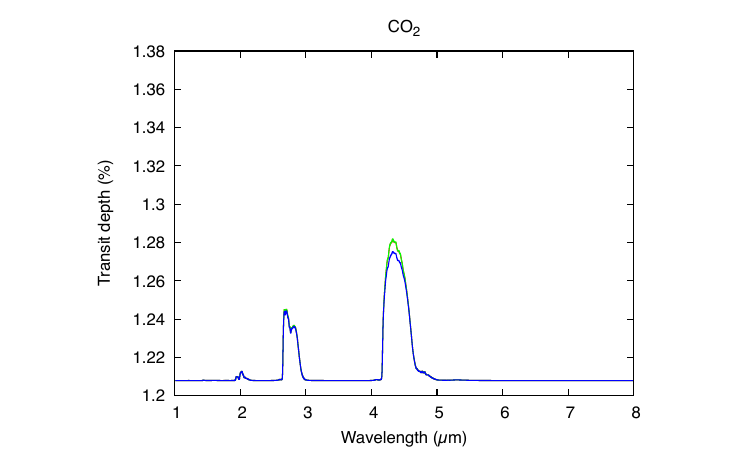}\\
\end{tabular}
\caption{Same as in Figure \ref{fig:abb1}, for the negatively reacting  species,  \ce{H2O}, \ce{CO}, and \ce{CO2}.} 
\label{fig:abb2}
\end{figure*}

Results are detailed in  Figures~\ref{fig:abb1} and \ref{fig:abb2}, where we show chemical profiles of groups of species  reacting positively (\ce{CH4}, \ce{C2H2}, \ce{HCN})  and not reacting (\ce{H2O},  \ce{CO}, \ce{CO2}) to the increase in $L_{\rm X}$. Together with their vertical profiles, we plot the corresponding individual contributions to the total transit depth. We also display the pressure $p$ at which the integrated contribution functions of individual species reach 10$\%$ and 90$\%$ of their maximum value. It is evident that the major contributions to the spectrum by \ce{CH4}, \ce{C2H2} and \ce{HCN} arise in those atmospheric layers where such species experience the largest abundance variations in response to the stellar forcing. Although depleted by the increase in the X-ray luminosity, water, \ce{CO}, and \ce{CO2} do not show significant spectral variations. 

\subsection{High resolution}
\begin{figure*} 
\begin{tabular}{cc} 
\hspace{-1cm}
\includegraphics[width=8cm]{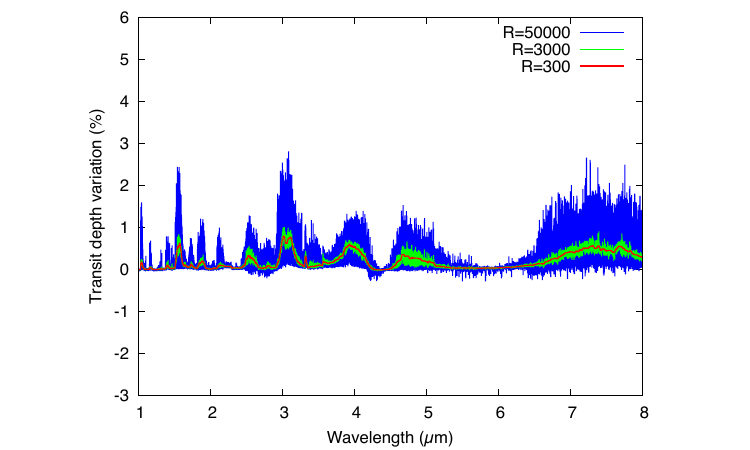} &  \includegraphics[width=8cm]{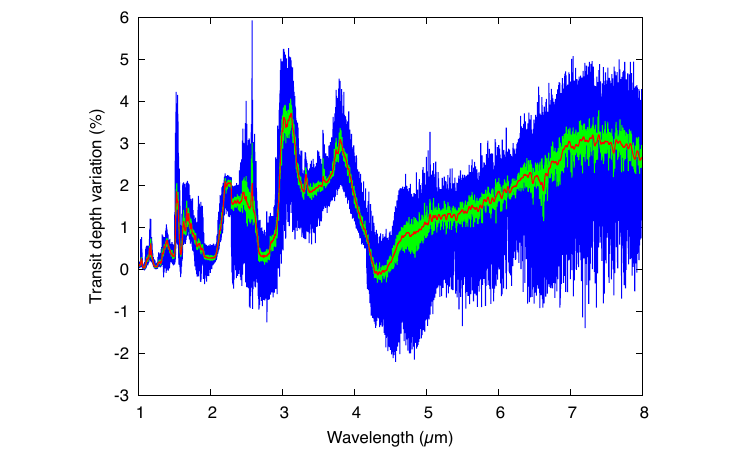} 
\end{tabular}
\caption{Difference spectra in percent between the model with $L_{\rm X} = 10^{28}$ and $L_{\rm X} = 10^{26}$~erg s$^{-1}$ (left panel), and the model with $L_{\rm X} = 10^{30}$ and $L_{\rm X} = 10^{26}$~erg s$^{-1}$ (right panel), constructed using three values of the resolving power, $R = 300$ (red line), 3,000 (green line), and 50,000 (blue line).}
\label{fig:highres}
\end{figure*}

In this Section we construct spectra using moderate ($R=3,000$) to high ($R=50,000$) resolving powers. We show the results in the form of difference spectra (in percent) between those induced by $L_{\rm X} = 10^{28}$ and $L_{\rm X} = 10^{30}$~erg s$^{-1}$ and $L_{\rm X} = 10^{26}$~erg s$^{-1}$ (Figure~\ref{fig:highres}). This last value of the XUV luminosity provides chemical effects impacting marginally on molecular abundances. 

From the results shown in Figure \ref{fig:highres} it is evident that high resolution spectra probe atmospheric layers at pressures lower than those sampled in the lowest resolution case, $R = 300$. As an example, variations in the \ce{CO} and \ce{CO2} profiles clearly arise from atmospheric layers located at pressure lower than $\sim 10^{-4}$~bar, not sampled by low resolution spectra (see below). 

\begin{figure*} 
\begin{tabular}{cc} 
\hspace{-1cm}
\includegraphics[width=8cm]{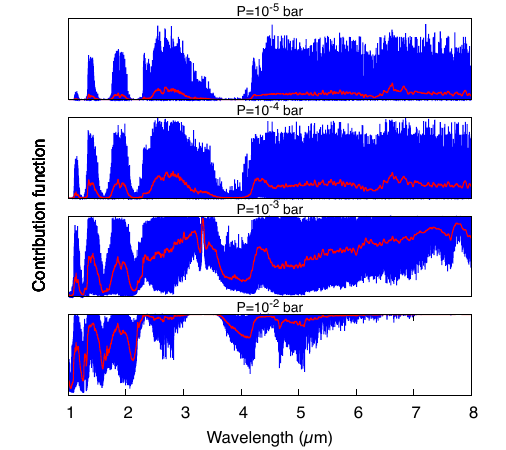} &  \includegraphics[width=8cm]{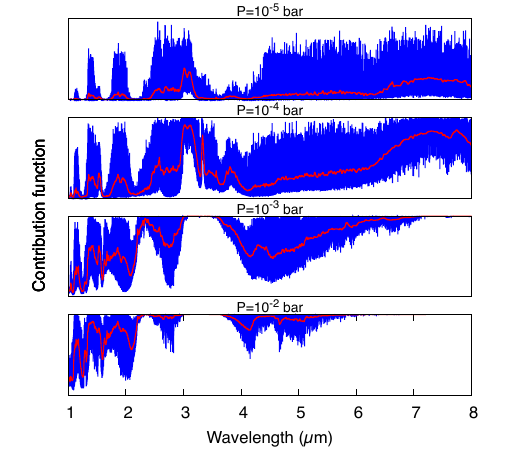} 
\end{tabular}
\caption{Wavelength-dependent contribution functions using low (red lines) and high (blue lines) resolution powers, for increasing values of the boundary pressure $P = 1 \times 10^{-5}$, $1 \times 10^{-4}$, $1 \times 10^{-3}$, and $1 \times 10^{-2}$~bar (top to bottom), for $L_{\rm X} = 10^{26}$~erg s$^{-1}$ (left panel) and $L_{\rm X} = 10^{30}$~erg s$^{-1}$ (right panel). We recall that contribution functions vary from zero (no signal) to 1 (full contribution).}
\label{fig:contrip}
\end{figure*}

The global impact on the spectra are summarized by the plot of the wavelength-dependent contribution functions, $c(\lambda,p > P)$ equation (\ref{conf}), in the low and high resolution cases, for increasing values of the boundary pressure $P$ as displayed in Figure \ref{fig:contrip}. 
The upper atmospheric layers are those where XUV irradiation affects more deeply the chemical profiles, so that suitable spectral features carry the imprinting of photochemical processes over the molecular abundances. For instance, a band around $4.5~\mu$m attributable to \ce{CO} and \ce{CO2} features, shows a decrease with increasing X-ray luminosity. This reflects the removal of these species via ionization (as discussed in \citealt{Locci22}), at pressures that are mostly inaccessible to low resolution observations. In the same spectral range is instead appreciable a positive variation in the transmission depth due to \ce{HCN}. This species also displays a strong feature at approximately 1.8~$\mu$m. Incidentally, the feature falls within the spectral range of the near infrared echelle spectrograph \emph{GIANO} \citep{oliva06} available at the Telescopio Nazionale Galileo (TNG), and it differs from its low activity counterpart by approximately 2.5\%. In the methane band at $3.3-4~\mu$m, the percentage variations sum up to 3$\%$, while in the $3-3.3~\mu$m and $7-8~\mu$m bands due to \ce{C2H2} and \ce{HCN}, the changes can reach up 5$\%$ in the case of high activity. Interestingly, spectral modifications arising in the high resolution case with $L_{\rm X} = 10^{28}$~erg s$^{-1}$ are of similar extent to those produced when $L_{\rm X} = 10^{30}$~erg s$^{-1}$ but observed in low resolution.

In general, high resolution observations sample outermost regions with pressures more than a factor of 10 lower than the corresponding observations performed at lower resolution. However, even if high resolution core lines form at high altitudes, profile wings experience pressures similar (if not larger) to low resolution profiles. 

\section{Discussion and conclusions}\label{sez:dis}
In this work, we investigate if possible disequilibrium processes in exoplanetary atmosphere, induced by photochemical activity may be detectable. This study is part of a more general debate on the combined application of  space-borne low resolution observations and ground-based high resolution spectroscopy (e.g., \citealt{Guilluy22}). 

Photochemistry affects carbon, nitrogen and oxygen species, although the dominant heavy molecular constituents, such as \ce{CO}, \ce{CO2}, \ce{H2O} and \ce{H2} are relatively stable against photochemical destruction or recycle efficiently. \ce{CH4} and \ce{NH3} are more interesting from a photochemical perspective, in particular when ionizing radiation is present \citep{Locci22}, and open the way to the formation of gas-phase hydrocarbons such as acetylene, \ce{C2H2} and ethylene, \ce{C2H4}. These species have been observed to play an important role in Solar System giant planets (e.g., \citealt{Sinclair19}). Moreover, polymerization of these initial photochemical products are likely and more complex hydrocarbons including e.g., tholins and soots may be expected to form. The present photochemical model predicts an enhancement of \ce{CH4}, \ce{C2H2} and \ce{HCN}, as the  the X-ray luminosity increases over $L_{\rm X} = 10^{28}$~erg s$^{-1}$, with synthetic transmission spectra presenting evident features arising from these species, even in the low resolution regime. Due to the large opacity difference between the core and the wings of molecular lines, high resolution spectral synthesis probes a broad range in atmospheric temperatures and pressures, with a predicted overall $3-4 ~\%$ variation in the spectra (with respect to stellar low activity). 

Space-borne low resolution spectra (e.g., those achievable with Ariel) will probe layers different from those observed by ground-based high resolution spectroscopy, so that their mutual contributions provide the opportunity to relate different regimes in a planetary atmosphere, such as clouds in the troposphere and line cores in the upper thermosphere (e.g., \citealt{Pino18}). For their very nature, molecules do not form and survive at extreme altitudes, residing at most at pressures around $p \sim 10^{-7}-10^{-8}$~bar. High resolution observations posses a dynamic range that seems able at constraining vertical variations in gas abundances in those regions. 
\begin{figure*} 
\begin{tabular}{cc} 
\hspace{-1cm}
\includegraphics[width=8cm]{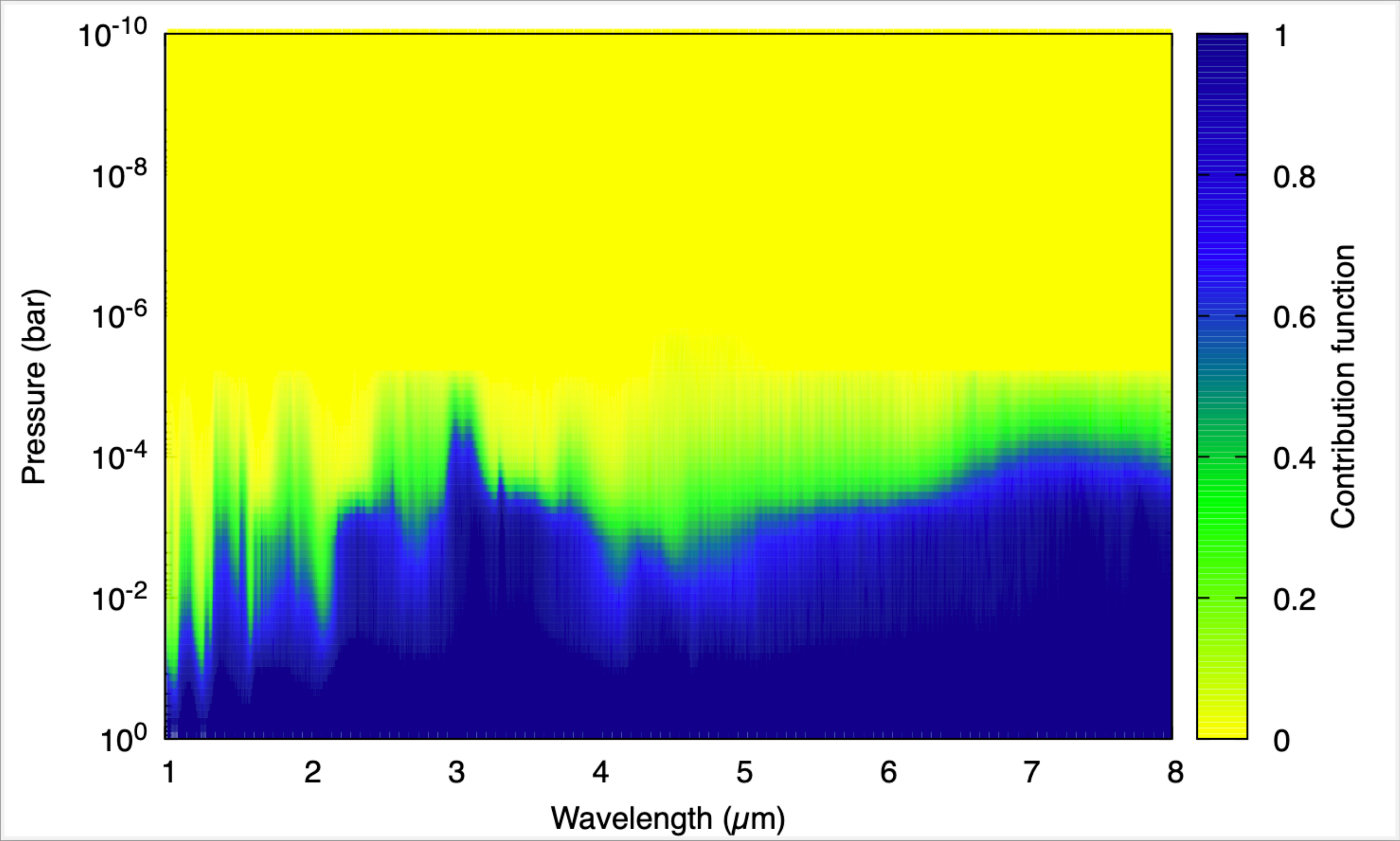} &  \includegraphics[width=8cm]{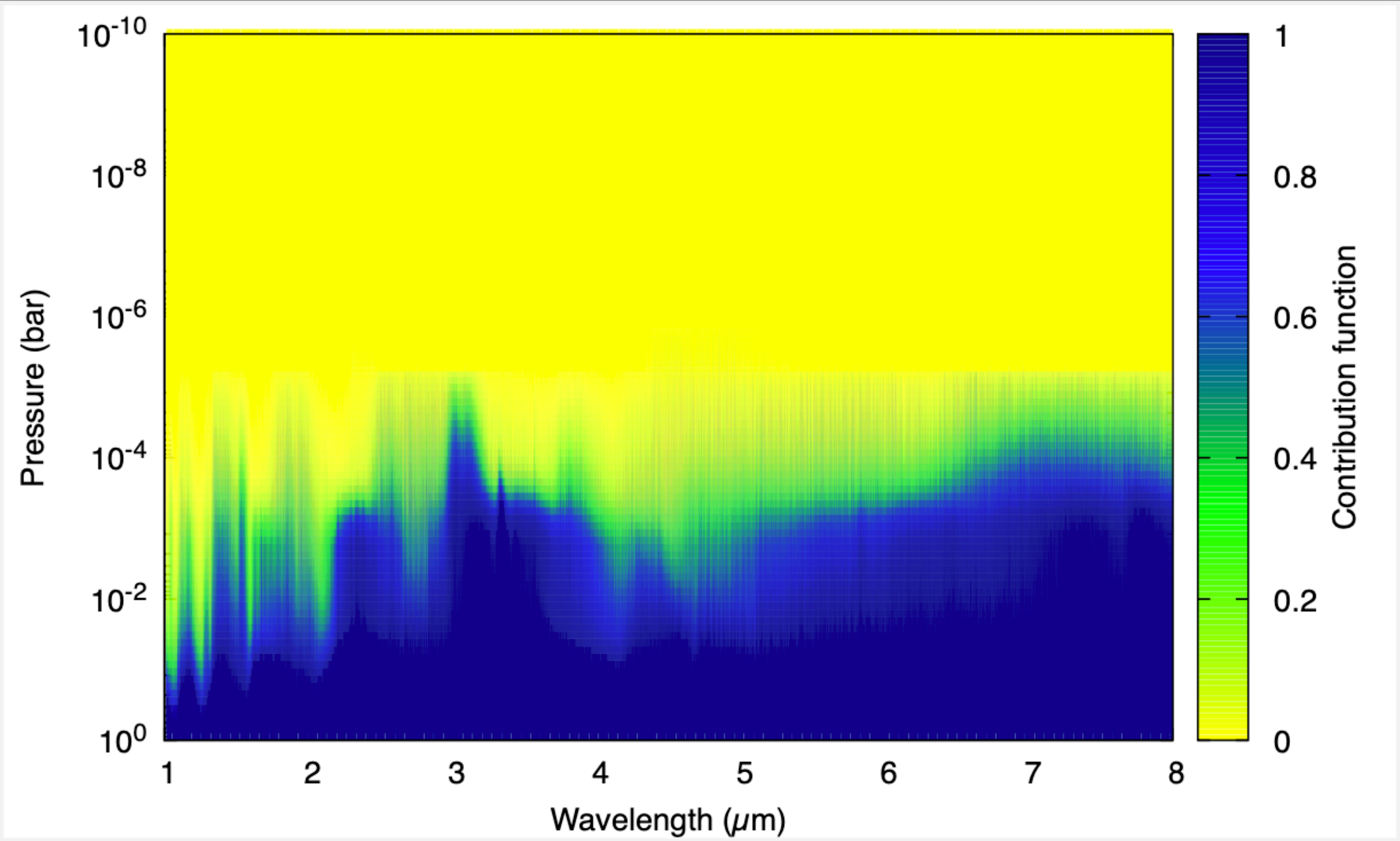} 
\end{tabular}
\caption{Wavelength-dependent contribution functions using low, $R = 300$, (left panel) and high, $R = 50000$, (right panel) resolution powers, displayed for the whole range of pressures (i.e., from $p = 10^{-10}$ to 1~bar). The X-ray intensity is $L_{\rm X} = 10^{30}$~erg s$^{-1}$.}
\label{fig:contrip2}
\end{figure*}
In Figure \ref{fig:contrip2}, we compare low $(R = 300)$ and high $(R = 50,000)$ resolution contribution functions  throughout  the vertical pressure profile.  In both cases, the atmospheric "bottom" for detections is located  at $p \sim 1 \times 10^{-2}$~bar (see also Figure \ref{fig:contrip}). The capability of high resolution observations in sampling lower pressures than those experienced in low resolution is evidenced by the significant population of the green region ($c \sim 0.2-0.4$)  by the highest (blue, $c \ga 0.6$) signals.   

\begin{table}
\caption{Molecular lines sensitive to X-ray radiation.}
\begin{tabular}{ccccc}
 & \ce{HCN}  &  \ce{C2H2}  & \ce{CH4} & \ce{CO}\\
 \hline
           1 &   \phantom{$^\dag$}3.0874$^\dag$     &   3.0528      &   3.1493      &   4.8336      \\
           2 &   3.0935     &   3.1171      &   3.2025       &   4.5732      \\
           3 &   2.9685     &   3.0288      &   3.2024       &   4.5132     \\
           4 &   3.0203     &   3.0205      &   3.1753       &   4.5456      \\
           5 &   7.5396     &   7.3136      &   3.2308       &   4.5389      \\
           6 &   7.0158     &   3.1098      &   3.2405       &   4.7736      \\
           7 &   3.0689     &   3.0090      &   3.1933       &   4.7359      \\
           8 &   3.1189     &   3.0264      &   3.1842       &   4.8337      \\
           9 &   7.5395     &   3.0389      &   3.3245       &   4.6332      \\
          10 &  7.4897     &   7.4855      &   3.1843       &   4.4664      \\
\hline
\end{tabular}
\label{Tab:Intenselines}
\flushleft 
$\dag$ Wavelengths in microns.
\end{table}

\begin{figure} 
\includegraphics[width=9cm]{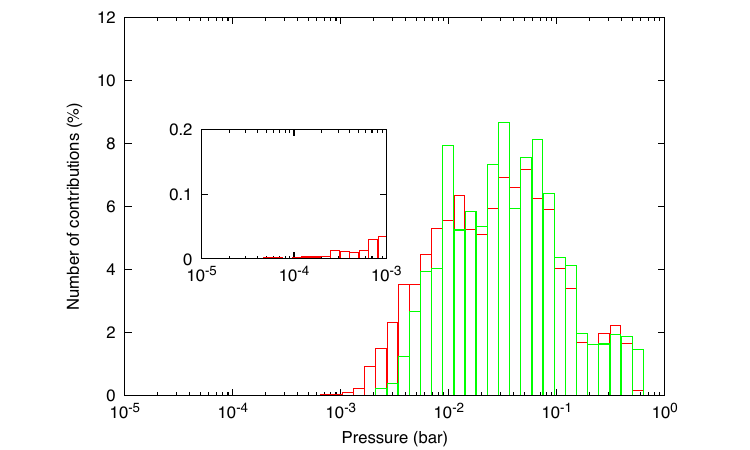} \\
\includegraphics[width=9cm]{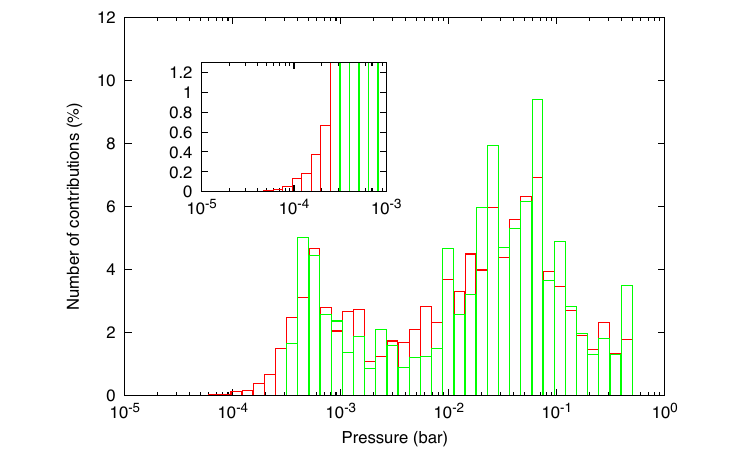}
\caption{Histogram of the number of lines arising from single pressure layers. Top panel: $L_{\rm X} = 10^{26}$~erg s$^{-1}$; bottom panel: $L_{\rm X} = 10^{30}$~erg s$^{-1}$. Green line: low resolution; red lines: high-resolution.}
\label{fig:isto}
\end{figure}

We may quantify the extra information contained in a high resolution spectral profile by listing, for each pressure layer, the number, wavelengths, and species of the transitions that contribute to the profile. This way, we may construct histograms like those reported in Figure~\ref{fig:isto}. In the high resolution case, a straightforward visual inspection reveals the existence of a tail of contributing lines (with $c \ga 0.9$) arising from pressures $p \la 3 \times 10^{-3}$~bar. With the exception of a small contribution in the high-activity case, low pressures are virtually excluded in low resolution profiles. 
In the high activity case, the number of contributions from upper atmospheric regions increases, with lines forming up to $p \sim 10^{-5}$ bar. Most of the lines arising from pressures as low as $p \lesssim 10^{-3}$~bar, belong to band systems of \ce{CO}, \ce{CH4}, \ce{C2H2}, and \ce{HCN}, species particularly sensitive to changes in the X-ray luminosity. A small set of the brightest molecular features is reported in Table \ref{Tab:Intenselines}. 

In addition to photo-chemistry other kinds of disequilibrium processes are expected to modify abundances within exoplanet atmospheres. Transport-induced quenching modify chemical equilibrium as a result of the dominance of transport processes like convection or large-scale eddy diffusion over chemical reactions. In our chemical code, we did not include dynamic prescriptions, including  zonal winds that tend to uniform the chemistry longitudinally to the dayside of the planet (e.g., \citealt{Agundez14}). Moreover, stars are variable in time, and they may be subjected to flares and other impulsive phenomena that can display sudden, drastic increases in brightness for a few minutes to a few hours. Such high activity may increase photochemical and ionization rates, that may impact atmospheric chemistry. Our aim is to assess the importance of photo-chemistry and the detectability of its effects imprinted either in low and high resolution transmission spectra. Through an accurate chemical model, we find that chemical profiles present trends in response to the interaction of the atmospheric gas with XUV radiation, that translate into evident spectral signatures, recognizable markers of the departure from equilibrium. The next step in our analysis will be to estimate the fate of these disequilibrium clues, i.e., if they will be preserved, destroyed or modified by coexisting and competitive kinetics-related processes. 

\section*{Acknowledgement}
We acknowledge contributions from ASI-INAF agreements 2021-5-HH.0 and 2018-16-HH.0. AM acknowledges partial support from PRIN INAF 2019 (HOT-ATMOS).  DL acknowledges contibutrions from  Bando di Ricerca Fondamentale INAF - MINI-GRANTS di RSN 2 and STILES (Strengthening the Italian leadership in ELT and SKA), Attività di progetto – IR 0000034, CUP C33C22000640006

\bibliography{spesynt.bib}{}
\bibliographystyle{aasjournal}
\end{document}